\documentclass[prb, twocolumn, showpacs]{revtex4}% Physical Review B

\usepackage{graphicx}% Include figure files
\usepackage{dcolumn}% Align table columns on decimal point
\usepackage{bm}% bold math

\begin{document}

\preprint{APS/123-QED}

\title{Observation of Mass Transport through Solid $^4$He}% Force line breaks with \\

\author{M.W. Ray}
\author{R.B. Hallock}%
\affiliation{%
Laboratory for Low Temperature Physics, Department of Physics,\\
University of Massachusetts, Amherst, MA 01003
}%

\date{February 10, 2009}% It is always \today, today,

\begin{abstract}
By use of a novel experimental design,  one that provides for
superfluid helium in contact with bulk hcp $^4$He off the melting
curve, we have observed the DC transport of mass through a cell
filled with solid $^4$He in the hcp region of the phase diagram.
Flow, which shows characteristics of a superflow, is seen to be
independent of the method used to grow the solid, but depends on
pressure and temperature.  The temperature dependence suggests the
possibility of hysteresis.
\end{abstract}

\maketitle

\section{Introduction}
\label{sec:intro} The possibility of a superfluid solid, or a
so-called\cite{Mullin1971,Matsuda1970} supersolid phase, in solid
$^4$He has been discussed for many
years\cite{Penrose1956,Andreev1969,Chester1970,Guyer1971,Meisel1992},
but until recently no evidence for such a state had been observed
experimentally. In 2004, motivated by the experiments of Ho,
Bindloss and Goodkind\cite{Ho1997}, Kim and Chan \cite{Kim2004a,
Kim2004b, Kim2005, Kim2006} reported a reduction in the resonant
period of a torsion oscillator filled with solid hcp helium.
Although their interpretation that this non classical rotational
inertia (NCRI), as it was called, was evidence of a supersolid
state remains controversial, it renewed an intense interest in the
subject. Since the Kim and Chan results, there have been several
other\cite{Penzev2007,Kondo2007,Rittner2007,Aoki2007} torsion
oscillator experiments performed confirming the earlier results,
but showing a considerable range of NCRI values. Rittner and Reppy
\cite{Rittner2007} observed a sample history dependence with large
``superfluid" fractions ($\sim$ 20 \%) measured in quench cooled
samples that had small macroscopic dimensions, and saw reductions
of the ``superfluid" fractions that depended on how much the
sample had annealed. Aoki \textit{et al.} \cite{Aoki2007} observed
a frequency-dependent NCRI and studied ``limiting velocities",
showing a complicated dependence on sample history. Day and
Beamish \cite{Day2007} studied the shear modulus of solid $^4$He,
and saw that it increased at approximately the same temperature as
the observed onset of substantial changes in the NCRI. This
surprising result was explained by the de-pinning of dislocations
imbedded in the solid. Because it showed a dependence on
temperature and $^3$He concentration similar to that in torsion
oscillator experiments, it was thought that the behavior of the
shear modulus might be closely related to the NCRI results.

Any superfluid phase should support a frictionless mass transport.
Therefore, if this non classical rotational inertia is indeed
evidence of supersolidity in helium, then a superflow should be
possible. Several experiments have investigated this directly, and
all found no evidence of such a transport of mass. Greywall saw no
flow when he attempted to push solid helium through 200 $\mu$m
diameter capillaries \cite{Greywall1977}.  A somewhat similar
experiment was performed recently by Day, Herman and
Beamish\cite{Day2005} in which they tried to push helium through
the small (7 nm) pores of Vycor glass. An experiment by Day and
Beamish\cite{Day2006} extended the results to glass capillary
arrays with 25 $\mu$m diameter pores. Recent work by Rittner et
al.\cite{Rittner2009} searched for flow in thin quench-cooled
samples.  No evidence for flow was seen in those experiments.
Bonfait \textit{et. al.} took a different approach
\cite{Bonfait1989}. They grew solid helium in a concentric,
cylindrical, U-tube-like geometry on the melting curve with liquid
helium on either side. They then attempted to observe
equilibration in the different levels of the solid free surface on
the two sides of the U-tube; no equilibration took place. A very
similar approach was taken by Sasaki \textit{et al.}
\cite{Sasaki2006, Sasaki2007} with the inclusion of a window to
allow direct visual images of the solid sample. The measurements
showed that the positions of the solid free surfaces shifted and
this was interpreted as being due to mass flow through the solid
along grain boundaries which were visually identifiable in the
sample \cite{Sasaki2006}; when grain boundaries disappeared, so
did the observable flow. Later analysis, however, showed that flow
was most probably occurring along liquid channels that exist where
a grain boundary meets the wall of the cell or where grain
boundaries intersect \cite{Sasaki2007,Sasaki2008}.

It is now believed theoretically that the observed NCRI behavior
of the solid helium is likely due to disorder and is not an
intrinsic property of a solid helium crystal. In fact it has been
shown theoretically that the original interpretations of the
torsion oscillator results as a Bose-Eintstein condensate of
vacancies in the crystal is likely wrong \cite{Boninsegni2006a},
and that a perfect vacancy-free solid helium crystal cannot become
a supersolid \cite{Prokofev2005,Ceperley2004,Clark2006}.
Simulations have shown that a frictionless flow of mass through
solid $^4$He can occur through defects in the solid, such as grain
boundaries \cite{Pollet2007} and dislocations
\cite{Boninsegni2007}. Simulations have also shown that under
certain conditions a glassy state \cite{Nussinov2007,Andreev2007}
of solid helium, a ``superglass"\cite{Biroli2008}, may allow for a
superflow \cite{Boninsegni2006}. However, whether or not these
defects can account for the observed NCRI has not yet been shown.
In particular, the understanding of superfluid fractions as large
as 20 \% in quench-cooled samples remains a mystery.

We have previously briefly reported on an experiment in which we
were able to induce mass flow through solid helium
\cite{Ray2008a,Ray2009}. This report provides additional data and
more extensive details of our work. Our approach is conceptually
different from the previous flow experiments of Greywall and
Beamish and his collaborators. In our approach we inject atoms
into the solid rather than applying a mechanical force to squeeze
the lattice. A key difference between our experiments and those of
Sasaki\textit{ et al.}\cite{Sasaki2006}, who observed a shift in
the relative positions of solid-liquid interfaces, is that we have
employed the properties of liquid helium in a confined geometry
(Vycor) to create a liquid-solid interface that is not on the bulk
melting curve. We are thus able to apply a chemical potential
difference directly across the solid without the application of
mechanical pressure to the lattice. This approach allows us to
perform zero-frequency flow experiments by injecting helium at low
temperatures and at pressures in excess of the bulk solidification
pressure, ($\sim$ 25 bar). In sec. \ref{sec:exp} we describe the
apparatus we have developed, and explain our procedures. In sec.
\ref{sec:grow} we discuss various methods we have employed for the
growth of our solid helium samples.  In sec. \ref{sec:res} we
present our results. Section \ref{sec:dis} discusses those
results, including various possible alternate explanations for our
observations. In sec. \ref{sec:con} we summarize our conclusions.
We also include appendices of atypical observations and include
tables, which provide details relevant to the data sets.

\section{Experimental Design}
\label{sec:exp}

\begin{figure}
\resizebox{3 in}{!}{
\includegraphics{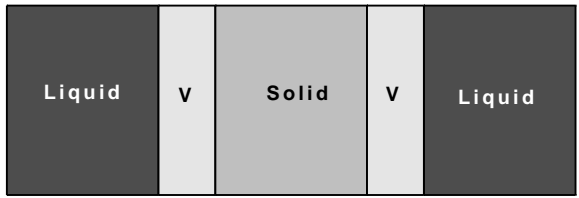}}
\caption{\label{fig:schem} Conceptual representation of the
experiment showing the sandwich-like geometry with three chambers,
each pair of which is separated by a region of liquid
helium-filled porous Vycor glass, V. In operation a temperature
difference is applied across each of the regions of Vycor, which
keeps the liquid reservoirs from freezing.}
\end{figure}
\begin{figure}
\resizebox{3 in}{!}{
\includegraphics{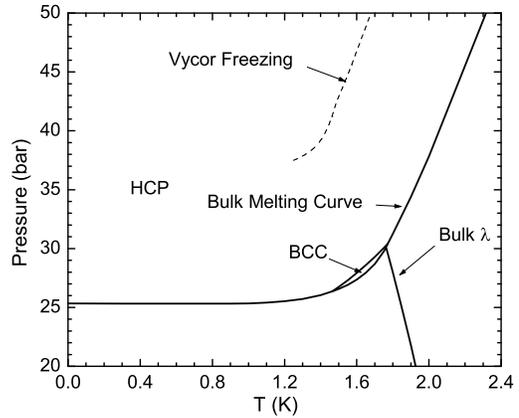}}
\caption{\label{fig:phase} Phase diagram of helium showing the
melting curve for helium inside of porous Vycor glass, adapted
from previous work \cite{Beamish1983,Lie-zhao1986,Adams1987}. }
\end{figure}
The concept of our experiment is very simple, and the basic idea
is shown in figure \ref{fig:schem}.  Three chambers are separated
from each other by porous Vycor glass.  The center chamber
contains the solid hcp helium sample while the outer chambers and
Vycor contain liquid helium. With the pressure of the outside
reservoirs, and hence in the Vycor, below $\sim$ 37 bar the helium
inside the Vycor will remain a liquid (figure \ref{fig:phase}) due
to the confined geometry provided by the pores
\cite{Beamish1983,Lie-zhao1986,Adams1987}. Thus, by imposing a
temperature gradient across the Vycor so that the liquid in the
reservoirs does not freeze, we can maintain liquid helium in
contact with solid helium with the bulk solid off the melting
curve at a given temperature.  To perform the experiment we simply
create a chemical potential difference between the two liquid
chambers by, say, injecting helium atoms into one side to raise
its pressure, a process we've termed a ``push" or ``injection."
Alternatively, we can lower the pressure on one side, which we
have termed a ``pull" or ``withdrawal." We then monitor the
pressure in the other fill line for a response; if the two
pressures come toward equilibrium then mass had to have moved
through the cell that contains solid $^4$He.

\begin{figure}
\resizebox{3 in}{!}{
\includegraphics{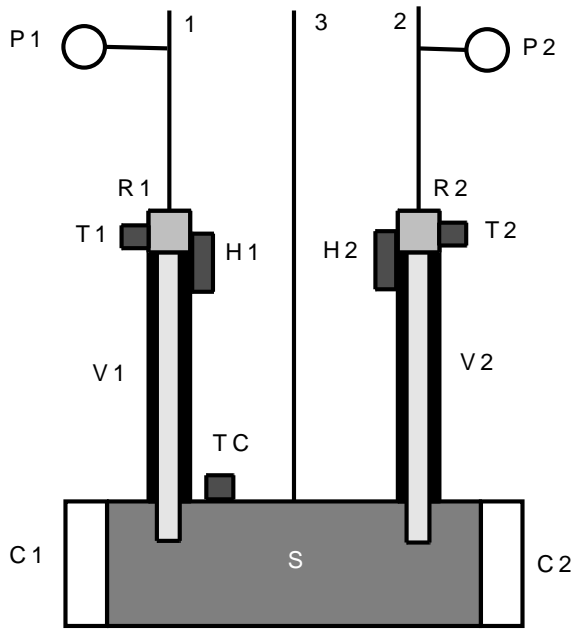}}
\caption{\label{fig:cell} Schematic diagram of the cell used for
flow experiments.  Three fill lines lead to the cell, two go to
liquid reservoirs R1 and R2 above the Vycor Rods V1 and V2.  The
third one leads directly to the solid chamber, S.  Two capacitance
pressure gauges, C1 and C2, sit on either side of the cell for
\textit{in situ} pressure measurements.  Pressures in the Vycor
lines (1 and 2) are read by pressure transducers outside of the
cryostat.  Each reservoir has a heater, H1 and H2, which prevents
the liquid in it from freezing, and the reservoir temperatures are
read by carbon resistance thermometers T1 and T2.  The cell
temperature is recorded by a third carbon resistance thermometer,
TC. The cell thermometer reading, denoted TC, provides the
temperature of the sample, T.}
\end{figure}
The cell designed for these flow experiments is shown
schematically in figure \ref{fig:cell}.  Three stainless steel
capillaries lead to the cell. Two of them (numbered 1 and 2) lead
to the liquid reservoirs, R1 and R2, which are at the top of the
Vycor rods, which have a diameter of 1.5 mm. These two capillaries
are heat sunk only at 4 K, which allows us to keep solid helium
from forming in them. Capillary 3, heat sunk at 1 K, leads
directly to the solid chamber S, which is cylindrical in shape
with a diameter of 6.35 mm. Capillary 3 was mainly used to aid in
the initial fill, and final evacuation of the cell, since under
some conditions, the flow of helium through the pores of the Vycor
was slow. The Vycor rods are epoxied into thin (0.8 mm wall
thickness) stainless steel tubes using Stycast 2850 FT. The
centers of the Vycor rods are positioned 20.6 mm apart; they
extend approximately 6 mm into S and each provides 0.30 cm$^2$ of
macroscopic surface area of contact between the Vycor and the bulk
solid in the sample chamber. Capacitance pressure gauges of the
Straty and Adams type \cite{Straty1969} are attached to each end
of the cell to measure the pressure of the solid \textit{in situ}.
This arrangement also allows us to measure any pressure gradients
in the solid that may appear, an issue that we will return to
later. There are two carbon resistance thermometers on each of S,
R1 and R2; one for temperature control and one for recording the
temperature (the temperature controlling thermometers are not
shown in figure \ref{fig:cell}). Two heaters, H1 and H2, were used
to maintain a temperature gradient and ensured that the liquid in
R1 and R2 did not freeze. The whole solid chamber, S, is bolted to
a copper plate which is attached to the mixing chamber of a
dilution refrigerator by six 6.35 mm diameter copper rods.

The heat flow down the Vycor rods was higher than we anticipated,
and is an issue that we hope to resolve with the next generation
of the apparatus. The stainless steel rods that house the Vycor
were designed to keep the heat flow from the warmer liquid
reservoirs to the cold solid chamber to a low level. We estimated
that with no helium in the Vycor, the cell at 100 mK and the Vycor
top at 1.7 K, the heat flow to the cell should be no more than
$\approx$ 20 $\mu$W.  When designing this experiment it was our
belief that liquid helium should not contribute much to the
thermal conductivity of the Vycor and stainless steel combination.
Ideally, the small pore size of the Vycor should act as a
superleak allowing superfluid to pass (which provides for a
fountain effect), but blocking the flow of normal fluid, thus
preventing thermal counterflow. Unfortunately the observed heat
load on our mixing chamber with helium in the apparatus was larger
than expected, and thus our lowest temperature with helium in the
cell was limited to $\sim 300$ mK, depending on the pressure.

This problem of the heat load could be attributed to one or more
causes. For instance, a hole through the Vycor that is several
times larger than the nominal pore size or a break in the epoxy
that gives normal fluid a path to move around the Vycor could
enhance the effective thermal conductivity. Subsequent to some of
our measurements we were advised that Vycor rods might have an
axial imperfection in which the Vycor structure differs from the
usual structure.  Evidence for this was first reported by Wilson,
Edwards and Tough \cite{Wilson1968}. Indeed, careful subsequent
inspection of several samples from our batch of Vycor rods
revealed a small axially symmetric imperfection ($\approx$ 25
$\mu$m in diameter) that appears to continue down the entire
length of the inspected rods. It is possible that this
imperfection is found only in Vycor rods \cite{BeamishPC}, but
plates may have a similar imperfection, but of planar geometry
\cite{Reppypc}. The elevated heat conduction could also be due to
fluid shorts along the side of the Vycor. For instance, the epoxy
might crack, possibly due to differing thermal expansion
coefficients, which might allow a path big enough for some limited
thermal counterflow to occur. The thermal expansion coefficient of
Stycast 2850 is 1.5 times greater than stainless
steel\cite{Pobell1996}; Vycor, if it is like many
glasses\cite{Pobell1996}, should have an expansion coefficient
smaller than stainless steel.

Despite the fact that the heat load has prevented us from reaching
the most desirable lower temperatures, it does not affect the
central conclusion of our experiment that we have observed a flux
of atoms through solid helium housed in our sample cell.  With
regard to the effect of the Vycor, any bulk liquid helium path
around or in the Vycor should freeze in the low temperature region
of the Vycor once the pressure goes much above the bulk melting
pressure. Although our current lowest temperature achieved is
higher than the temperature at which substantial changes in NCRI
set in for the torsion oscillator experiments ($\sim$ 60 - 80 mK),
we have nonetheless been able to grow solid helium, learn about
the behavior of this system, and make interesting observations.

\section{Sample Growth}
\label{sec:grow}
\begin{figure}
\resizebox{3 in}{!}{
\includegraphics{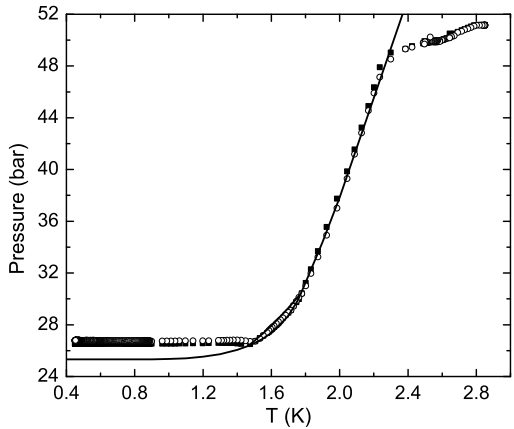}}
\caption{\label{fig:bc} Solid growth by the blocked capillary
technique.  The two sets of symbols represent data from the two
pressure gauges on the cell, C1, C2, and the solid line is the
bulk-helium melting curve.}
\end{figure}
There are several different techniques that can be used to grow
solid helium crystals.  Each one is likely to create solid $^4$He
samples of different quality, with differing numbers of defects.
Techniques that minimize stresses as the crystals grow are
expected to result in higher quality samples. Since the torsion
oscillator results are now thought to be caused by the disorder in
the solid, it would be interesting to study a variety of growth
techniques in our experiments.

The most widely used growth technique is the blocked capillary
method. In this method a plug of solid helium is formed in the
fill line, and the cell is cooled with a constant number of atoms
present. During solid helium growth the pressure and temperature
change substantially as the growing solid follows the melting
curve, which causes strain in the solid. It is for this reason
that blocked capillary growth is thought to produce more
disordered crystals. Furthermore, crystals grown this way may pass
through the bcc region of the phase diagram, which is also
believed to adversely effect the crystal quality\cite{Mikhin2007}.
Our cell allows us to grow crystals using a slightly modified
blocked capillary technique. The 1 K heat sink on line 3 allows us
to form a solid helium plug in that capillary. As we will
describe, we could limit the liquid from entering the cell from
lines 1 and 2 since while the cell is on the melting curve the
lines are filled entirely with normal fluid because we control the
pressure in the fill lines.

Figure \ref{fig:bc} shows a typical trajectory of the cell for
blocked capillary growth. Starting near 51 bar, and 2.8 K we
freeze the helium in line 3, and start cooling the cell. The
intersection with the melting curve, $P_{melt}$, is marked by a
sharp change in $dP/dT$. The trajectory then follows the melting
curve until the growth of the solid has finished at which point
the cell is filled with bulk solid. While the phase trajectory of
the helium in the cell moves down the melting curve, we slowly
lower the pressure in fill lines 1 and 2 to closely match the
pressure of the cell in a process designed to mimic a fully
blocked capillary and limit the migration of atoms into or out of
the cell. A modest number of atoms entering the cell through the
Vycor should not strongly modify the amount of disorder in the
solid which is likely dominated by the substantial pressure and
temperature changes encountered as the solid follows the melting
curve.

\begin{figure}
\resizebox{3 in}{!}{
\includegraphics{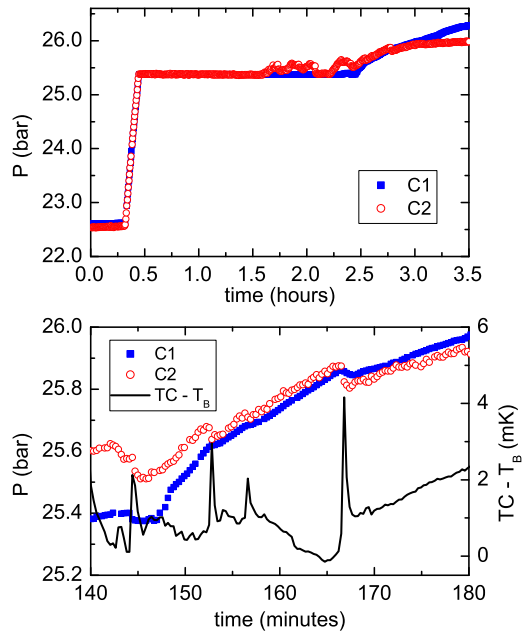}}
\caption{\label{fig:spikes2} (color online) An example of the data
from the capacitors C1 and C2 during the growth of a solid sample
(sample AB) from the superfluid at fixed temperature.  Top: A
record of the growth of solid hcp He$^4$ from the superfluid
showing the pressure of the solid in S during growth as measured
on capacitors C1 and C2 during an injection of atoms though lines
1 and 2 beginning at t $\approx$ 0.3 hours. Bottom: An enlarged
view of the cell pressure as atoms are fed into S via both V1 and
V2. A slowly changing background drift of the cell temperature,
TC, was observed and an exponential {\it vs.} time (T$_B$) was fit
to the data and was point-wise subtracted from TC to enhance
visibility of the temperature transients seen on the cell
thermometer TC for the cell at TC = 415 mK. Transient temperature
increases accompany the downward steps in the pressure recorded on
C2 and also weakly visible on C1.}
\end{figure}

A second approach is to grow the solid from the superfluid at
constant temperature. A typical trajectory for this type of growth
is shown in figure \ref{fig:spikes2} (top figure) for the growth
of sample AB (Table I, Appendix A). In this method we add atoms to
the cell through capillaries 1 and 2 (which lead to the Vycor).
 As the pressure in the cell increases it eventually hits the melting
curve, and as long as there is still liquid in the cell, the
pressure stays constant, and we continue to feed atoms into the
cell through lines 1 and 2. Eventually the cell is filled with
solid and the pressure can continue to increase. Since those lines
do not freeze we can continuously add atoms even after the cell
has crossed into the bulk solid region.

An interesting observation, which can be seen in the vicinity of t
= 1.5 - 2.5 hours in the top section of figure \ref{fig:spikes2},
is the presence of pressure drops that appear in the cell pressure
in the vicinity of the melting curve. These pressure drops are
accompanied by sharp transient rises in the cell temperature as
shown in the lower section of figure \ref{fig:spikes2}. It is
believed that these anomalous events are due to either a
re-organization of the crystal, or meta-stable liquid regions
freezing and we have discussed them at greater length
separately\cite{Ray2009a}.

Growing solid helium from the superfluid at constant temperature
is thought to produce higher quality crystals as long as the cell
pressure $P = P_{melt}$ since the liquid is superfluid and the
pressure is evenly distributed throughout the sample. However, as
soon as the pressure rises above the equilibrium melting pressure,
the further addition of atoms produces a large amount of stress
which most likely results in a highly disordered solid. This is
supported by visual observations of solid helium grown this way
that document crystals that are cloudy in
appearance\cite{Ford2007, Sasaki2007}, an indication of a highly
disordered solid helium sample.

Another growth technique, which is utilized by the community less
frequently, is growth at constant pressure. In this method, the
cell is cooled while atoms are continuously fed to the cell
through the capillaries 1 and 2 to keep the pressure constant. We
speculate that because the growth and subsequent cooling will
occur at constant pressure, the stresses in the crystal will be
smaller, and thus this approach should produce higher quality
crystals. None of our experiments to date have utilized this
method.

  Almost all
of the samples grown in torsion oscillators have been grown with
the blocked capillary method. Bonfait \textit{et al.}
\cite{Bonfait1989}, and Sasaki \textit{et al.} \cite{Sasaki2006}
have grown crystals from the superfluid, however they were limited
to performing their measurements on the melting curve. It is not
yet well understood how crystal growth should effect the behavior
recently seen in solid helium.  Given the spread in NCRI values
reported in the various torsional oscillator experiments, even in
ones with cells of similar geometry, it is clear that differences
among samples result in substantial differences in the
quantitative results.

\section{Results}
\label{sec:res}
\subsection{Liquid Flow Through Vycor}
\begin{figure}
\resizebox{3 in}{!}{
\includegraphics{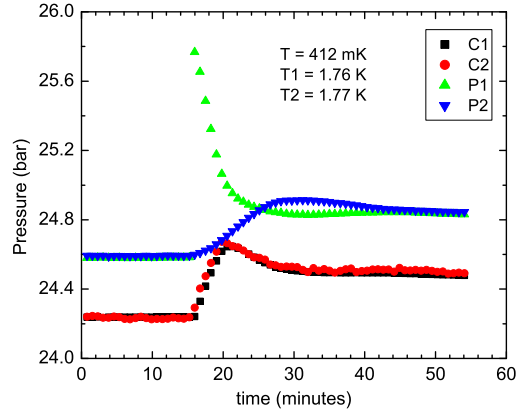}}
\caption{\label{fig:lhe} (color online) An injection of atoms into
line 1 with liquid helium in S began at $t$ $\approx$ 16  min.
Here $T1$ and $T2$ are the temperatures at the liquid reservoirs
at the top end of the Vycor rods and TC is the temperature of the
cell. P1 and P2 are the pressures at room temperature in fill
lines 1 and 2, and C1 and C2 are the capacitive pressure readings
at either end of the sample cell. Here, in the region where the
rate of change of P2 is linear, dP2/dt $\approx$ 1.67 mbar/sec.
The difference between the pressure read on the capacitors, C1 and
C2, and those read on the gauges P1 and P2, $\approx$ 0.4 bar can
be accounted for by the fountain effect.}
\end{figure}

\begin{figure}
\resizebox{3 in}{!}{
\includegraphics{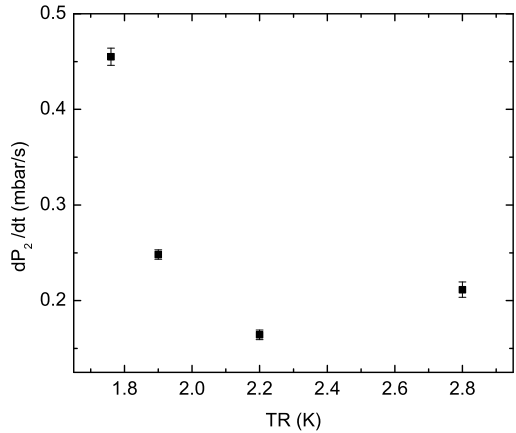}}
\caption{\label{fig:vflow}Rate of pressure change measured in line
2 following an increase in the pressure in line 1 for various
reservoir temperatures, TR = T1 $\approx$ T2, for cell pressures
in the range 23.4 to 23.8 bar.}
\end{figure}

In order to determine the properties of the flow through the Vycor
regions contained in V1 and V2, we typically fill region S with
liquid helium and measure the flow rate under varying conditions
similar to those that we expect to encounter with solid helium in
S. Figure \ref{fig:lhe} shows the results of one of these studies.
In this case T1 = 1.76 K, T2 = 1.77 K and TC = 412 mK,
temperatures that are similar to those present when there is solid
in S. There is an offset between the cell pressure and the fill
line pressure which can be accounted for by the fountain effect.
The presence of this fountain effect with liquid helium in the
apparatus, consistent with expectations, is suggestive that any
parallel liquid flow paths in or around the Vycor are rather
minor. At t = 16 minutes after the data record started, helium was
injected into line 1 to create a pressure difference between the
two lines. As seen, there is a subsequent increase in the cell
pressure and line 2 pressure as atoms flow through the Vycor with
equilibrium being achieved at about t = 30 minutes. For a span of
time P2 changes linearly with time, which should be expected for
superflow at a critical velocity. Figure \ref{fig:vflow} shows the
time rate of change of the pressure, dP2/dt, in the regime where
the pressure changes linearly, for various reservoir temperatures,
T1 $\approx$ T2. As expected the rate decreases as the temperature
increases because more normal fluid is in the Vycor at higher
temperatures. Note that the upper regions of the Vycor near the
reservoirs are outside the superfliud region of the phase diagram
for helium in Vycor. Therefore, some pressure dependence of the
flow through the Vycor was anticipated.

Our observed flows through solid helium are a small fraction of
the observed limiting flows though Vycor with liquid in the cell.
Since our liquid-only flow measurements (of the sort shown in
Figure 6) may be influenced by possible parallel flow channels, we
are unable to conclude conclusively that the flows we observe with
solid in the cell are unaffected by the flux though the Vycor.
While we believe that such effects due to the Vycor are not large
(and will comment on this quantitatively later) we recognize that
caution is necessary in a discussion of flow rate limits.

\subsection{Solid Helium}

\begin{figure}
\resizebox{3 in}{!}{
\includegraphics{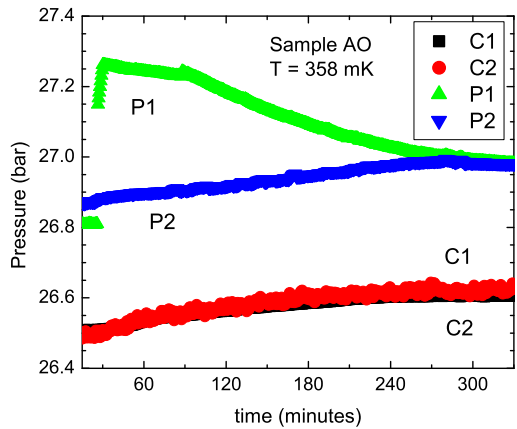}}
\caption{\label{fig:figureAO} (color online) Sample AO, created
from superfluid at 358 mK, showed a flow of mass through solid
helium. The pressure in R1, P1, was increased at t $\approx$ 30
minutes, the regulator feeding helium to line 1 was closed at t
$\approx$ 90 minutes, and changes in pressure were observed for
about 5 hours. Note that dP2/dT was nearly linear for a
substantial duration and independent of P1-P2. Note C1 = C2. (For
additional comments on C1 and C2, see Section V, part A:
discussion of flow and pressure gradients.) }
\end{figure}

\begin{figure}
\resizebox{3 in}{!}{
\includegraphics{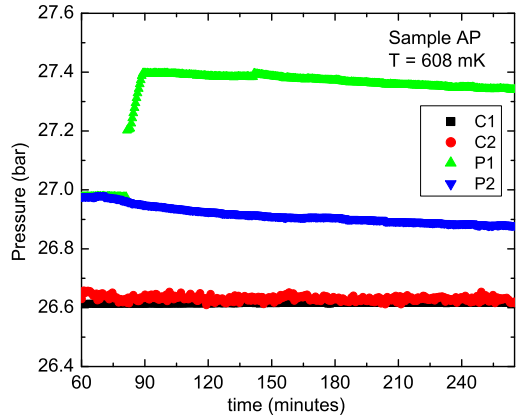}}
\caption{\label{fig:figureAP} (color online) Sample AP, created
from sample AO by warming to 608mK, showed no significant flow of
mass through the solid helium.  Note C1 = C2 $\approx$ constant. }
\end{figure}

After creating a solid in region S, we attempt to document flow
through the solid. To illustrate the behavior we will show several
examples. The first example is that for a sample (AO, Appendix A,
Table I) grown from the superfluid and studied at 359 mK (figure
\ref{fig:figureAO}). After growth, the sample sat idle for
$\approx$ 18 hours.  For our first measurement on this sample (we
call such first-measurement samples ``freshly made samples") the
detailed data record is started and at t = 30 minutes the pressure
to line 1 is increased by 0.45 bar and the regulator that feeds
atoms into line 1 is closed at t = 90 minutes, which terminates
the feed. Note that P2 shows a nearly linear increase with time
and the rate of change of P2 with time is independent of P1 - P2.
This is a very typical example of data that we interpret as
evidence for the flow of atoms through the cell filled with solid
helium, in this case from line 1 to line 2. The only way for P2 to
have increased was for atoms to have moved from R1 to R2 by travel
through the solid helium in region S.  Hence we can conclude from
figure \ref{fig:figureAO} that mass has flowed through the cell
filled with bulk solid helium.  Note that the capacitors each
record a change in pressure of the solid in the cell.

Following this measurement, the sample was warmed to 608 mK and
designated sample AP. (Since this was the second measurement on
sample AO, we renamed the sample AP and do not refer to AP as a
fresh sample.)  An attempt to observe flow was then made by
application of an increase in pressure to line 1 by 0.42 bar. We
interpret the result (figure \ref{fig:figureAP}) as evidence for
no flow.  We have observed this behavior on a number of occasions,
including one case where P1 - P2 = constant $\approx$ 0.5 bar for
nearly 20 hours.

\begin{figure}
\resizebox{3 in}{!}{
\includegraphics{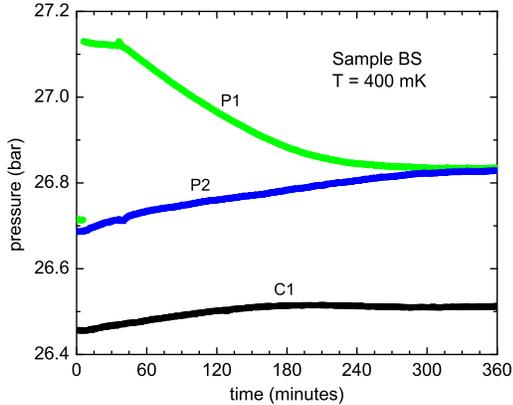}}
\caption{\label{fig:flow} (color online) Sample BS showing a flow
of mass through solid Helium.  The pressure in R1, P1, was
increased at t $\approx$ 6 minutes, the regulator feeding helium
to line 1 was closed at t $\approx$ 30 minutes, and changes  in
pressure were observed for about 6 hours.  Note that dP2/dT was
nearly linear for a substantial duration and independent of P1-P2.
}
\end{figure}

Figure \ref{fig:flow} shows another data set that provides an
example of evidence for flow through a sample of solid helium
(sample BS) that was grown from the superfluid. At t $\approx$ 6
minutes P1 was abruptly increased by about 0.4 bar, and atoms were
continuously fed into the reservoir via line 1. After t $\approx$
30 minutes the regulator was shut off and the system was allowed
to evolve. We observed an equilibration in P1 and P2 after about 5
hours, an increase of the pressure in S recorded by C1
(measurements from C2 were not available at the time this data set
was taken) and a nearly linear increase in P2 while P1 fell.

Although the flow rates observed depend on temperature and
pressure, the behavior seen in figures \ref{fig:figureAO} and
\ref{fig:flow} is typical of all cases where we have seen flow. It
is interesting to note that in both cases C1, the pressure in S,
increased indicating that \textit{some} of the atoms made their
way into the solid but did not move up V2 to the other capillary.
This sequence of events is seen every time flow is observed
through a sample.  That is, upon injection of atoms into line 1,
the pressure in the cell and the pressure in line 2 go up in
similar fashion to figures \ref{fig:figureAO} and \ref{fig:flow}.
One might be led to conclude from the changes in C1 and C2 that
plastic flow\cite{Suzuki1973} might be present in the solid
helium. We doubt that this is the case because, as we will
describe below, we see no such changes in C1 or C2 for cases of
similar cell pressure, but higher temperature, where we see no
evidence for flow.

\begin{figure}
\resizebox{3 in}{!}{
\includegraphics{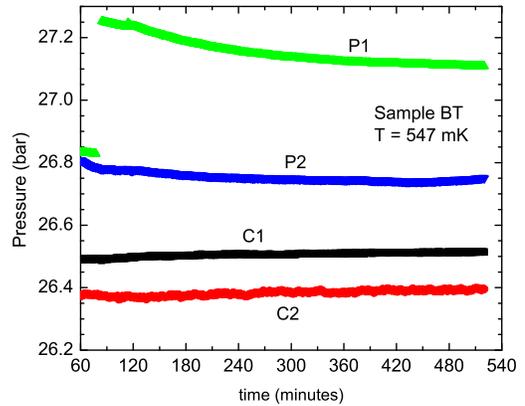}}
\caption{\label{fig:noflow} (color online) Sample BT was warmed
from 400 to 547 mK from sample BS. This is an example of a sample
that did not show flow. In this case the regulator fed atoms into
R1 for about 30 minutes, but over 7 hours there was no significant
movement of the pressures towards equilibrium. Note C1 $\neq$ C2.
(see section V, A) }
\end{figure}

Figure \ref{fig:noflow} presents another example of a sample (BT,
which was warmed from BS) that did not show evidence for flow.  In
this case, P1 was increased by $\approx$ 0.422 bar, and atoms were
fed in for 30 minutes. Although P1 did fall for a few hours, after
more than 7 hours there was no significant movement in the
pressures towards equilibrium. Thus we can conclude that no flow
was present through this sample.

A significant common difference that is seen between samples that
showed flow and those that did not is that in the latter no change
in the pressure recorded on C1 and C2 is measured. A number of
measurements similar to samples AO/AP and BS/BT have been made.
The measurements, sample histories, conditions under which the
measurements were made, and flow rates for all of the samples that
we have studied are summarized in table \ref{tab:samp} in appendix
B.

\begin{figure}
\resizebox{3 in}{!}{
\includegraphics{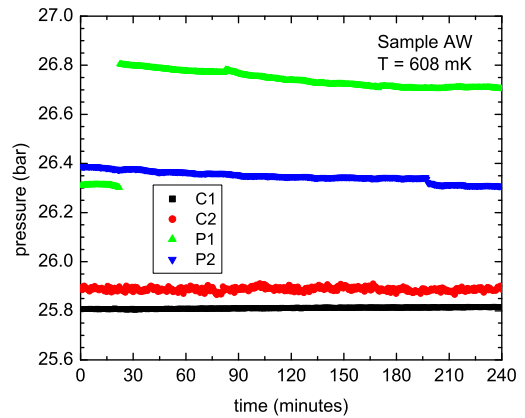}}
\caption{\label{fig:figureAW} (color online) Sample AW, which was
created fresh at 608 mK.  We interpret this as an example that did
not show long term evidence for flow. }
\end{figure}

\begin{figure}
\resizebox{3 in}{!}{
\includegraphics{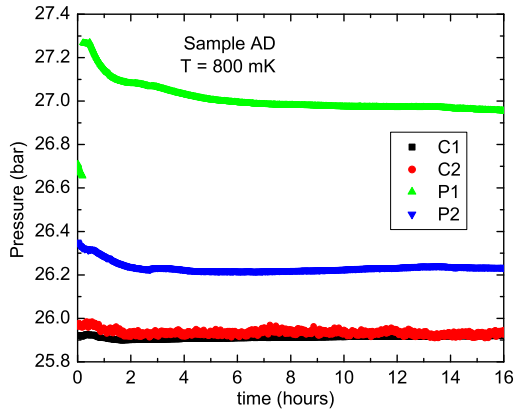}}
\caption{\label{fig:figureAD} (color online) Sample AD, which was
created fresh at 800 mK; an example of a sample that did not show
evidence for long term flow. }
\end{figure}

Samples can, of course, be created at higher temperatures.  Figure
\ref{fig:figureAW} is an example of a sample created at 600 mK. We
interpret the results of the application of a pressure increase to
line 1 of 500 mbar as evidence for no flow.   Another example is
shown in figure \ref{fig:figureAD} where a sample created at 800
mK showed no evidence for flow. This behavior is typical for all
samples we have created at or above 550 mK (or warmed to 550 mK or
above); none showed evidence for flow following, at times, initial
slow transient behavior.

\begin{figure}
\resizebox{3 in}{!}{
\includegraphics{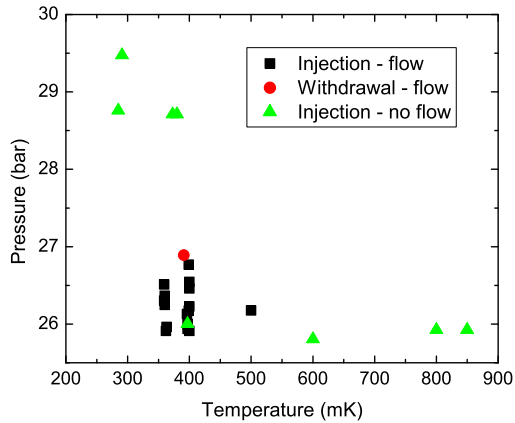}}
\caption{\label{fig:freshphase} (color online) Summary of the
results of flow attempts on freshly made samples listed in
appendix B, table I.  For fresh samples, the data fall into rather
clear regions of the phase diagram that show flow or no flow. The
technique to create a particular sample is tabulated in table I.
The highest pressure samples were gown by the blocked capillary
technique. }
\end{figure}

In figure \ref{fig:freshphase} we show the phase diagram
coordinates for most of our samples that were freshly made and the
results of efforts to observe flow. The data fall into two rather
clear regions: samples made at a lower pressure and temperature
show flow, those made at higher temperatures or pressures do not.

\subsection{Thermal Cycles}

Several of the samples listed in appendix B, table \ref{tab:samp},
are part of three-step sequences (call the steps a, b, c) in
which, after growing the solid helium, we then added atoms to line
1 and measured dP2$_a$/dt at a cell Pressure of P$_a$ and
temperature T$_a$. The temperature was then changed to some new
value, T$_b$ with pressure P$_b$ (usually higher due to the first
addition at T$_a$), for another addition to line 1, and dP2$_b$/dt
was measured at this new temperature. Finally, we lowered the
sample temperature to T$_c$ which is approximately the original
temperature and with pressure, P$_c$ which again was usually a bit
higher.  A third addition to line 1, produced the measured
dP2$_c$/dt.

\begin{figure}
\resizebox{3 in}{!}{
\includegraphics{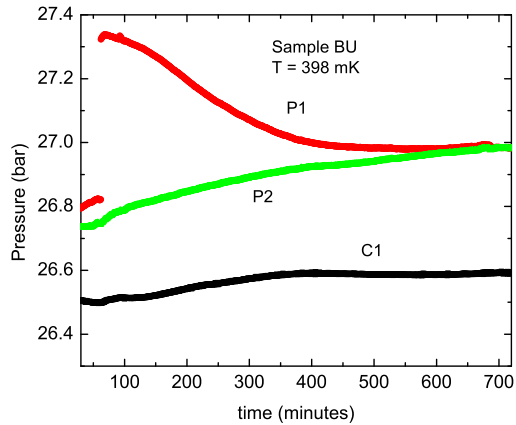}}
\caption{\label{fig:figureBU} (color online) Sample BU, which was
cooled from to 547 mK from sample BT and returned to a temperature
similar to that of sample BS.  The flow seen in sample BU was
similar to that seen in sample BS.  The measurements made with C2
were not well behaved for this data set and are not shown. }
\end{figure}

We have done several of these sequences where we cycle the
temperature as described above, and the results are shown in
appendix B, table \ref{tab:cycle}. The previously discussed
freshly-grown sample, BS was part of one such sequence (BS-BT-BU).
Figure \ref{fig:figureBU} shows the result of the third step, step
c, in that sequence; BT was cooled back to the temperature of BS,
renamed BU, and the flow rate, which had disappeared for sample
BT, returned to nearly the value that was present for sample BS.
The flow rate thus appears to have a temperature dependence.

In another example of a temperature cycle sequence (series 3 in
table \ref{tab:cycle}), we grew a sample at T = 498 mK (P = 26.37
bar), then increased P1 by an addition of atoms and observed flow
into line 2 (sample AK). The sample was then cooled to 360 mK
(sample AL), and P1 increased with flow again observed into line
2, this time at a slightly faster rate then the first time.
Finally the sample was warmed back to 498 mK (sample AM), and at a
still higher pressure of 26.71 bar.  A final injection of atoms to
this sample produced a behavior that was reminiscent of both flow
and no flow; the pressure, P1 decreased after shutting off the
regulator while C1 and C2 both increased; however, the pressure in
line 2 did not rise implying that there was actually no helium
flow through V2.

This sequence was then repeated with a new sample This sample was
grown cold then warmed (series 6, BD, T = 390 mK, P = 25.86 bar)
). In this case, flow was observed at 400 mK, but not at 500 mK,
though at 500 mK, again, C1 and C2 increased with P1 decreasing.
When the sample was cooled back to 400 mK, the addition of helium
to line 1 produced no flow into line 2 for ~60 minutes after which
an rather fast flow was observed.

The observations of the behavior of the flow for a sample warmed
to 500 mK, coupled with the absence of flow at 550 mK and above,
lead us to believe that at the pressure at which the sample was
studied perhaps there is a transition between a flow state and a
non-flow state around T = 500 mK.  We thus did two more thermal
cycles this time warming the sample to 450 mK. The first, series 7
(Samples BJ, BK, BL), showed the same behavior as the 500 mK
samples with P1 decreasing, C2, C1 increasing and P2 staying
constant, then again the same behavior when cooled back to 400 mK.
The second cycle (series 8, samples BP, BQ, BR)), however, showed
a clear flow and relaxation at 450 mK and then again when cooled
back to 400 mK. As previously described, the final sequence,
series 9, BS-BT-BU, was warmed to 547 mK and showed no flow, then
when cooled back to 400 mK, the flow returned. Measured near the
untimely end of our data run, this is the only clear example to
date for which we've seen a return to a nearly equal flow rate,
following the observation of no flow at higher temperatures,
without first performing a withdrawal.

\section{Discussion}
\label{sec:dis}
\subsection{Flow and Pressure Gradients}
Several conclusions can be immediately drawn from the data shown
in table \ref{tab:samp}.  First, the method of solid growth seems
to have no effect on whether or not flow is observed.  Several
samples grown from the superfluid have shown flow (for instance
sample A) and several have not (sample H). Likewise we have
observed flow in some samples grown by the blocked capillary
method (sample V) and we have also made samples by the blocked
capillary method that have not shown flow (sample D). We find it
curious, however, that samples created by the blocked capillary
technique and cooled at approximately constant pressure once off
the melting curve can show flow, while samples created from the
superfluid at or warmed to 800 mK do not show flow when
subsequently cooled.  Differences in cooling rates or defects in
the various samples may be involved.

While most of the runs listed in table \ref{tab:samp} were done by
injecting atoms into reservoir 1, one of them (sample U) was an
injection into reservoir 2, and it showed flow. Additionally, we
have several times attempted a subtraction of pressure in
reservoir 1 (a procedure we've termed a ``pull" or a
``withdrawal"). In each instance a withdrawal produced a flow of
atoms from the opposite reservoir (i.e. a ``withdrawal" from
reservoir 1 produced a corresponding drop in pressure in reservoir
2).  It is possible to induce a flow of mass in either direction.

As mentioned previously, the flow rates seen in figures
\ref{fig:figureAO} and \ref{fig:flow} are constant in time and
independent of the pressure difference between the two fill lines;
$dP2/dt \approx$ constant. This is reminiscent of a superflow
flowing at a limiting velocity, where the flow rate should be
independent of the pressure head.  Here it is useful to return to
the possible influence of the Vycor on the measured flow rates
through the solid helium.  If the Vycor were indeed causing an
upper limit to the flow (as opposed to the the solid itself), then
fresh samples grown at the same temperature and pressure should
show the same flow rate. We have a few fresh samples that were
grown at the same temperature and pressure. Samples A and AB are
one example of such a pair; both were fresh at 26.75 bar and 398
mK. For sample A, dP/dt = 0.0051 mbar/s, a much slower flow than
AB, for which dP/dt = 0.0230 mbar/s, even though A had slightly
lower Vycor temperatures and a bigger initial pressure step.
Another example is present in samples AN and AR, which were both
grown at 359 mK. AN was at a pressure of 26.30 bar and had a flow
of 0.0203 mbar/s. AR was at 26.25 bar, nearly the same pressure,
with a flow rate of 0.0088 mbar/s. Both had similar Vycor
temperatures, and similar pressure steps.  Certainly the flux
through the Vycor did not limit the lower flow rates observed for
the two sample pairs. Examples such as this provide evidence that
we are indeed observing critical flow limitation in the solid,
limitation that is sample dependent and not Vycor dependent. It
would be useful to repeat observations of this sort under the same
conditions with varying initial pressure increases to see if that
effects the flow rate. This is difficult to accomplish in practice
because the pressure of the cell changes with each injection, and
separately prepared samples may have different flow properties
(due to different configurations or different numbers of flow
paths).

\begin{figure}
\resizebox{3 in}{!}{
\includegraphics{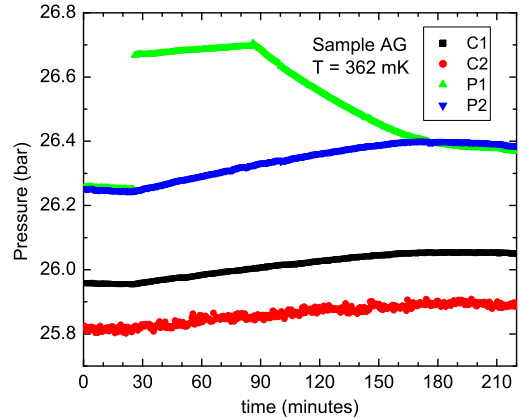}}
\caption{\label{fig:figureAG} (color online) Sample AG, created
from superfluid and studied at 362 mK, showed a flow of mass
through solid helium. Note in this case that C1 and C2 were not
equal; (compare to figure \ref{fig:figureAO}). }
\end{figure}

We find that the pressures recorded by C1 and C2 on opposites ends
of the cell are often different.  $\delta$C = C1 - C2 can
sometimes differ by several tenths of a bar and remain stable
(independent of the presence or absence of a flow), which means
that the solid lattice can support and maintain a pressure
gradient across the solid (which provides further evidence against
the presence of plastic flow). These pressure gradients tend to
occur after the solid is off the melting curve, but while we are
still adding atoms to it during the growth process. Sometimes we
see a relaxation of the pressure during a measurement, but often
times these pressure gradients persist until the solid is melted.
The presence of pressure gradients sustained in the solid seem to
have no effect on the flow. We have seen flow occur when C1 = C2,
i.e. $\delta$C = 0 (e.g. figure \ref{fig:figureAO}) and also when
C1 $\neq$ C2, i.e. $\delta$C $\neq$ 0 (e.g. figure
\ref{fig:figureAG}), which implies that the flow is not directly
related to pressure gradients within the solid.  As stated
perviously, when no flow is observed we also observe no change in
C1 and C2; with C1 = C2 (e.g. figure \ref{fig:figureAP}), or with
C1 $\neq$ C2 (e.g. figure \ref{fig:noflow}). We believe the reason
for this is that in order for the C1 and C2 to record a pressure
change, mass must move from the liquid in the Vycor to the
pressure gauges located on the ends of the solid. Whatever
mechanism is transporting the mass from V1 to V2 is likely also
responsible for moving the mass to the ends of the solid. With the
conducting paths no longer able to support mass flow, there is no
way to move mass to the ends of the cell where the pressure gauges
are, since we know from Day and Beamish \cite{Day2006} and Day
Herman and Beamish \cite{Day2005} that squeezing the lattice
cannot produce any mass flow. It should therefore come as no
surprise that we see no pressure change in the solid when no flow
is observed. These observations further appear to make unlikely
plastic flow\cite{Suzuki1973} as a flow mechanism, since it should
occur even if any conducting paths in the solid are unable to
conduct mass.

\subsection{Possible Hysteresis}

The data we have collected to date point to a temperature
dependence with possible hysteretic behavior.  All of the samples
created from superfluid at, or cycled to, T $\geq 550$ mK showed
no flow, even if they showed flow at lower temperatures, and
further, no flow is typically observed after cooling a sample from
T$\geq 550$ mK; i.e., a sample that was previously grown at or
warmed to T $\geq 550$ mK from a lower temperature. These
thermally cycled samples, however, could be made to flow again by
applying a withdrawal, and subsequent injections after such a
withdrawal produced flow through the sample. But, we have also
performed a withdrawal on a sample created and measured at 600 mK
which showed no evidence of flow. This is important because
samples at lower temperatures almost always flowed during a
withdrawal.  We can thus conclude that solid helium created from
superfluid at, or warmed to T$\geq 600$ mK does not support flow.
Furthermore, samples warmed to and then cooled from T$\geq 600$ mK
also do not flow without first withdrawing atoms from one fill
line. On the other hand, samples warmed to ~500 mK tend to show
some elements of flow, but mass does not flow into line 2. It
appears as if at these temperatures the flow paths are on the edge
of a transition from a flow state to a no flow state (or absence
of continuity of the conducting path across the sample). The data
thus point to a scenario in which whatever conducts the flow
ceases to do so for T$\geq 550$ mK.

Our data have also led us to believe that the liquid channels
responsible for the results of Sasaki \textit{et
al.}\cite{Sasaki2007} are likely \textit{not} responsible for the
flow observed in our samples. Liquid channels, as described by
Sasaki \textit{et al.}, should be superfluid at temperatures well
above 1K, which is contrary to what we have observed.  All samples
warmed to, or created at temperatures $>$550 mK show no flow by
either ``injection" or ``withdrawal" meaning that flow ceases in
whatever is conducting the mass flow through the solid (or the
conducting pathway disappears) at these temperatures. Annealing is
unlikely since we are well outside the temperature range where
annealing takes place on the timescales of our measurements, and
we see none of the effects of annealing that other labs see, such
as a decrease in the pressure of the sample
\cite{Mikhin2007,Rittner2007} that are seen when samples are close
the melting curve. We think it more likely that we are seeing flow
along structures, such as dislocations and grain boundaries
imbedded in the solid, which are predicted to be superfluid
\cite{Pollet2007, Boninsegni2007}.  In fact, Pollet \textit{et
al.} predict that grain boundaries should become superfluid for T
$\sim$ 500 mK; a prediction perhaps supported by our data that
shows flow that stops at and above 550 mK. In almost all cases
when we have warmed samples to temperatures above 550 mK, we do
not see flow when the sample is cooled. This could point to a
hysteresis in the flow behavior (i.e. flow might once again appear
if we were able to get the sample colder). Although we cannot yet
unambiguously rule out mass flow along liquid channels as an
explanation for our observations, we feel the data point more
strongly to flow along defects, although the matter is far from
completely settled\cite{Balibar2008a, Ray2008e}.

\subsection{Quantitative Aspects}

The data from Figure \ref{fig:flow} (sample BS) and other data
like it can be used to characterize the flow, and to make
quantitative comments relevant to what may be causing the flow.
The mass flux of a superfluid with density $\rho$ through a
conduit of cross section A and velocity $v$ can be written as
\begin{equation}
\frac{dm}{dt} = \rho v A.
 \label{eq:flow}
\end{equation}
For sample BS, using the time that we fed atoms into line 1, 30
minutes, and the rate at which the pressure in P1 falls
immediately after ceasing the addition of atoms, we can estimate
that we supplied as an upper limit an amount of mass
$\Delta$m$_{1}\approx $ 1.1$\times 10^{-4}$g.  Of this mass,
$\approx$ 6.6$\times 10^{-5}$g joined the solid, while $\approx$
4.6$\times 10^{-5}$g flowed through the solid and into line 2.
Thus, the average rate that mass flowed into line 2 can be
estimated as, dm$_2$/dt $\approx$ 2.2$\times 10^{-9}$g/sec.
Ignoring possible effects due to the Vycor, if this flow is along
dislocations pervasive throughout the solid as predicted by
Boninsegni \textit{et al.}\cite{Boninsegni2007}, then we can take
as our conducting pathways tubes with a diameter of atomic
dimensions, ~0.5 nm.  If we assume that the critical velocity is
on the order of that in a thin helium film, then we can take $v
\sim 200$ cm/s. Putting these numbers into equation \ref{eq:flow},
and using $\rho \approx$ 0.19 g/cm$^3$ as the density gives a mass
flow through one dislocation of $\approx$ $~7.5 \times
10^{-14}$g/s, and thus it would take something on the order of
$2.9 \times 10^{4}$ dislocations to account for the mass flow we
observe in figure \ref{fig:flow}.  We can compute similar numbers
for the other measurements from table \ref{tab:samp}. If we take
measurements done on freshly prepared samples at T = 400 $\pm$ 4
mK, and P = 26.1 $\pm$ 0.2 bar we see the number of dislocations
needed to support the flow range in number from $\approx$ 2$\times
10^4$ to $\approx$ 5$\times 10^4$.  Using the cross section area
of S between the Vycor rods, 0.3 cm$^2$ we can compute the
dislocation density for these samples to be in the range of 6
$\times 10^4$ cm$^{-2}$ to 16 $\times 10^4$ cm$^{-2}$.

We can do a similar analysis assuming that the conducting defects
are grain boundaries instead of dislocations.  If our solid sample
contains one grain boundary that spans the entire diameter of the
cell and is one atomic layer thick, then the cross sectional area
of the conducting path is A = $3.2 \times 10^{-8}$ cm$^2$.  Then,
using equation \ref{eq:flow} and taking account of the fact that
the intersection with the Vycor is a length less than the diameter
of the cell, we can compute the velocity of the mass flowing
through this pathway as $v \approx 1$ cm/s.  Instead, if we were
to take the velocity of flow along the gain boundary to be 200
cm/sec, then a grain boundary of atomic thickness would need be
only 0.001 cm in width; i.e. the grain boundary only spans a
portion of the cell, which seems unrealistic.

We must also explore quantitatively the possibility of liquid
channels proposed by Sasaki et. al \cite{Sasaki2007, Sasaki2008},
which were discussed earlier.  They suggest that the size of the
liquid channels depends on 1/$\Delta P_{eq}^2$ where $\Delta
P_{eq}$ is the difference between the solid pressure and the
solid-liquid equilibrium pressure. If we adopt the view that
liquid channels are indeed present, then following Sasaki et. al
\cite{Sasaki2007, Sasaki2008} we can write the cross sectional
area of the channel, A$_{LC}$ as
\begin{equation}
A_{LC} =  R^2 \left (2 \sqrt{3} \sin (\phi) \sin(\phi +
\frac{\pi}{3}) - 3 \phi \right ) \label{eq:lca}
\end{equation}
Here $\phi =  \pi / 6 - \theta$ where $\theta  \approx \pi / 12$
is the  contact angle between the grain boundaries, and R is the
radius of curvature between the liquid and solid phases, which is
given by
\begin{equation}
R =  \frac{\rho_S}{\rho_S - \rho_L}\frac{\sigma_{LS}}{\Delta
P_{eq}} \label{eq:lcrad}
\end{equation}
where $\rho_S = 0.19$ g/cm$^3$ and $\rho_L = 0.17$ g/cm$^3$ are
the solid and liquid densities, and $\sigma_{LS} = 1.7 \times
10^{-4}$ N/m is the liquid-solid surface tension.

\begin{figure}
\resizebox{3 in}{!}{
\includegraphics{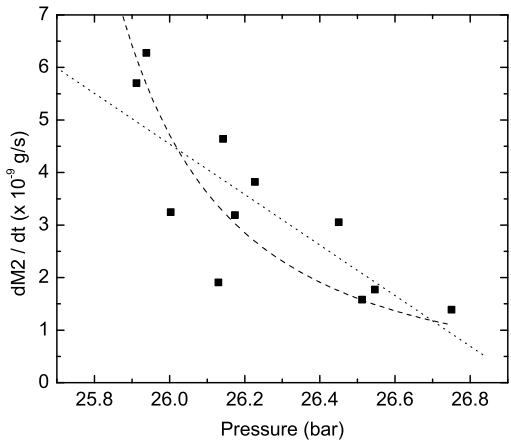}}
\caption{\label{fig:flowp2} Rate of change of mass in reservoir 2
after injecting atoms into reservoir 1 as a function of initial
sample pressure. All samples portrayed here were freshly grown in
the vicinity of 400 mK.  The curved line is a fit to the data
under the assumption that liquid channels carry the flow (equation
\ref{eq:lca}).  The straight line is an arbitrary straight line
fit.  The quality of the fit is nearly the same in each case.}
\end{figure}

To show the observed effect of cell pressure in our experiments,
we have plotted in figure \ref{fig:flowp2} the mass flow rate into
capillary 2 as a function of starting sample pressure for all of
our freshly made samples created at a temperature of T $\approx$
400 mK. Although it is difficult to make comparisons about flow
rates of separately prepared solid samples, since the number of
conducting pathways or configurations can be different for each
sample, it is clear that there is a general trend of decreasing
flow rate for increasing sample pressure. Figure \ref{fig:flowp2}
suggests that at T $\approx$ 400 mK there might be a sample cell
pressure above which the mass flux is $\approx$ zero.  It is
possible that there is a temperature-dependent critical pressure,
P$_H$(T), such that at temperature T no flow is observed above
that pressure. If this is the case then it means that for TC
$\approx$ 550 mK, P$_H$ $\approx$ P$_{melt}$.

Equation \ref{eq:flow} with equation \ref{eq:lca} as the cross
sectional area of the flow path is also plotted in figure
\ref{fig:flowp2} as the curved line with a limiting velocity taken
to be 800 cm/sec\cite{Harrison1974} and the number, N, of liquid
channels used as a fitting parameter. The fit results in N = 57
liquid channels spanning the distance between the Vycor Rods,
which should be interpreted as an average number for the set of
samples included. The fit is reasonable, but there is much scatter
in the data, and a linear fit (also shown) works equally well.  We
take the result as suggestive, but our observations that no flow
is ever observed at temperatures greater than 550 mK remain
unexplained by the liquid channel scenario.  As we have pointed
out\cite{Ray2008e}, presuming such channels remain present they
should continue to conduct above 550 mK, and be present for
samples made fresh at and above this temperature. But, we have not
been able to observed flows for T $\geq 550$ mK. Thus, we believe
that the weight of the evidence does not favor liquid channels as
the source for our observations.

Finally, we might assume that the mass is conducted not along
defects, but through the actual solid such as in a superglass
phase that has been predicted to occur in highly disordered
samples \cite{Boninsegni2006,Nussinov2007,Biroli2008}. In this
case, we can modify equation \ref{eq:flow} slightly to include a
dimensionless term $\xi$ which represents the fraction of the
solid that is in flux (if this were a superfluid, $\xi$ would
denote the superfluid fraction) so that
\begin{equation}
\label{eq:flowfrac}
\frac{dm}{dt} = \xi \rho v A
\end{equation}
In this case A is the cross sectional area of S between the Vycor
rods. (An alternate perspective is to assume that A is the open
cross sectional area of the pores in the Vycor, where they meet
the solid.  Below, numbers in parentheses are based on this
alternate perspective.) For the data in figure \ref{fig:flow} we
find that $v\xi = 2.9 \times 10^{-9}$ cm/sec ($v\xi = 1.2 \times
10^{-8}$ cm/sec). If we take $\xi \approx$ 0.01 (as is typical of
a number of the NCRI measurements), then $v \approx$ 2.9 $\times$
10$^{-7}$ cm/sec ($v \approx$ 1.2 $\times$ 10$^{-6}$ cm/sec). If
instead we arbitrarily take $v = 100 \mu$/sec, then $\xi = 2.9
\times 10^{-6}$ ($\xi = 1.2 \times 10^{-5}$). For the freshly made
samples at 400 mK and 26.1 bar, with $v = 100 \mu$/sec, we find
$\xi$ to be between 3 $\times 10^{-6}$ and 8 $\times 10^{-6}$ (1.2
$\times 10^{-5}$ and 3.2 $\times 10^{-5}$). These numbers are, of
course, highly dependent on the arbitrary estimate we used for the
critical velocity, but we can note that $\xi$ is several orders of
magnitude less than the superfluid fraction reported by Kim and
Chan in their bulk solid measurements of NCRI \cite{Kim2004b}.

\subsection{Chemical Potential}

\begin{figure}
\resizebox{3 in}{!}{
\includegraphics{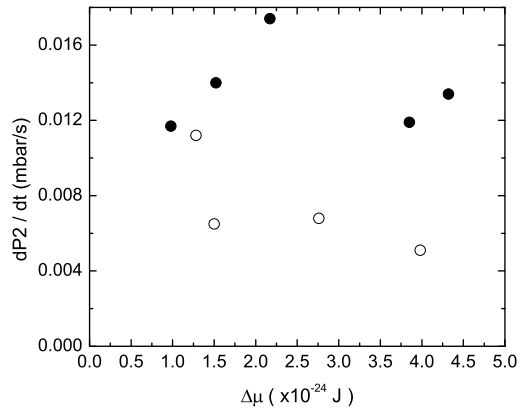}}
\caption{\label{fig:chem} Flow rate into line 2 versus the applied
chemical potential difference across the solid for fresh samples
at T $\approx$ 400 mK. Closed circles are at P = 26.5 $\pm$ 0.1
bar and open circles are at P = 26.1 $\pm$ 0.1 bar.}
\end{figure}

Since, as we have stated before, we are inducing flow by creating
a chemical potential difference across the solid, it is perhaps
useful to use the pressure and temperature data to calculate the
chemical potential difference applied across the solid. The
chemical potential, $\mu$, can be found from
\begin{equation}
\label{eq:chem} \mu (P,T) = \int \frac{V}{N} dP - \int \frac{S}{N} dT
\end{equation}
where $V$ is the volume, $S$ is the entropy and $N$ the number of atoms.
Using N = $\rho$V/m$_4$ where m$_4$ is the mass of a helium atom, and
defining the specific entropy s as the entropy per unit mass, s = S/$\rho$V,
equation \ref{eq:chem} becomes
\begin{equation}
\label{eq:chem2}
\mu(P,T) = m_4 \left( \int \frac{dP}{\rho} - \int s dT \right)
\end{equation}
It is sufficient in our case to calculate the chemical potential
of only the liquid above the Vycor.  In the Vycor there should be
no chemical potential difference since the flow is likely below
critical velocity (as we can tell from our measurements of the
flow of atoms through our system, e.g. figure \ref{fig:vflow}). So
by computing $\Delta \mu$ bewteen the tops of the Vcyor rods,
which are at a temperature of T$\approx$1.8 K, we can find the
chemical potential difference that is driving the flow through the
solid. Using equation \ref{eq:chem2}, we compute the chemical
potential for each side, then take the difference, $\Delta \mu =
\mu_1 - \mu_2$. Figure \ref{fig:chem} shows the measured flow rate
into line 2 {\it vs.} the applied chemical potential difference
across the solid at $T = 396 \pm 4$mK and for pressures of 26.5
$\pm$ 0.1 bar and 26.1 $\pm$ 0.1 bar. Although there is
considerable scatter, it appears that the flow rate into line 2 is
independent of the chemical potential difference between the two
lines, which provides further evidence of superflow at critical
velocity.  It also seems that the flow rate in samples at the
higher pressure of 26.5 $\pm$ 0.1 bar is lower than the flow rate
for the lower pressure samples regardless of the applied $\Delta
\mu$. Although, as mentioned before, there is no reason why two
separately prepared samples should have the same flow rate given
the same parameters (sample pressure, $\Delta \mu$, etc.), figure
\ref{fig:chem} does seem to show that there may be dependence of
the flow rate on the sample pressure.

\section{Summary}
\label{sec:con}

We have performed experiments in which a chemical potential
difference is applied across hcp solid $^4$He at low densities by
injecting liquid helium into one side of the solid, and have
observed a D.C. mass flow. The flow is observed at temperatures
below approximately 550 mK, and at pressures below approximately
26.9 bar. The flow rate is mostly constant over time, and it is
independent of the applied chemical potential difference which
leads us to believe that we are seeing a superflow at close to
critical velocity, mindful of our earlier caution about the
possible effects of the Vycor. The flow rate is dependent on the
pressure of the solid with flow substantially reduced by 27 bar at
400 mK. It is our thought that this pressure dependence of the
flow is itself a function of temperature, and as we move to lower
temperatures in future work the maximum pressure at which we see
flow may increase; indeed, in more recent work we have seen clear
evidence for flow at 120 mK and 28 bar.  We have also observed
that samples thermally cycled to, or above, 550 mK do not support
flow again when cooled down without first subtracting pressure
from one of the fill lines. This behavior could suggest
hysteresis, and in order to restore flow without a withdrawal of
pressure, we may have to get to lower temperatures. This behavior
could also be caused by defects introduced into the solid by the
process of withdrawal.

We conjecture that, based on the current evidence, the flow is
being conducted along defects in the solid, such as has been
predicted theoretically for grain boundaries and dislocations.  We
do not believe that the flow is along the liquid channels shown to
exist by Sasaki \textit{et al.} for the primary reason that at 550
mK, the temperature at which we cease to see flow, these liquid
channels should still be superfluid. Even if these liquid channels
did cease to conduct flow at such temperatures, presuming that
they remain in place, there is then no reason why flow shouldn't
return upon cooling the sample again. There also seems to be no
reason why they should not be present in fresh samples made for T
$\geq 600$ mK.

Finally, the relationship between our experiment and other solid
helium experiments, mainly the torsion oscillator experiments and
the shear modulus experiments, is yet to be determined, and the
only way to concretely establish such a relationship will be to
extend our results to lower temperatures.  It is possible that, as
shown by our much lower ``superfluid" fraction, that we are seeing
some precursor effects of the mechanism that is causing the NCRI
in the torsion oscillators, which is not visible to them due to
the very small effective NCRI fraction implied by our
measurements. Indeed, many torsional oscillator experiments begin
to see evidence for period shifts in the vicinity of 250 - 300 mK.
With enhanced sensitivity it is possible that they would see
evidence for NCRI at higher temperatures.  This makes measurements
at lower temperature of the utmost importance.

We would like to thank B. Svistunov and N. Prokofev for
discussions that motivated this experiment.  We also thank S.
Balibar and J. Beamish for helpful advice on the growth of solid
helium and for many productive conversations.  Also, F. Caupin,
M.C.W. Chan, R.A. Guyer, E. Kim, H. Kojima, M.W. Meisel, W.J.
Mullin, J.D. Reppy, S. Rittner, E. Rudavaskii and Ye. Vekohov
provided many useful comments and discussions. This work supported
by NSF DMR 06-50092 and CDRF 2853 and by access to the facilities
provided by the MRSEC at the University of Massachusetts Amherst
supported by NSF DMR 02-13695 and DMR 08-20506.

\appendix

\section{Unusual Observations}

\begin{figure}
\resizebox{3 in}{!}{
\includegraphics{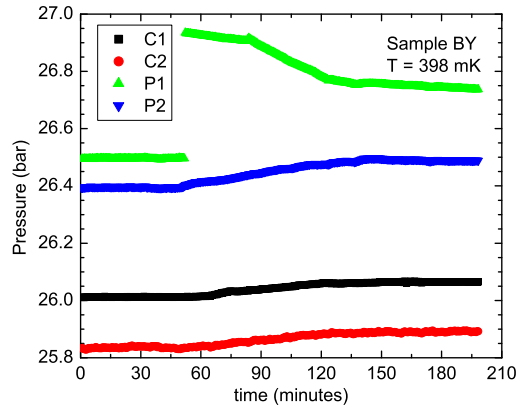}}
\caption{\label{fig:figureBY} (color online) Sample BY, which
showed the unusual behavior of a flow that stopped before
equilibration. }
\end{figure}

\begin{figure}
\resizebox{3 in}{!}{
\includegraphics{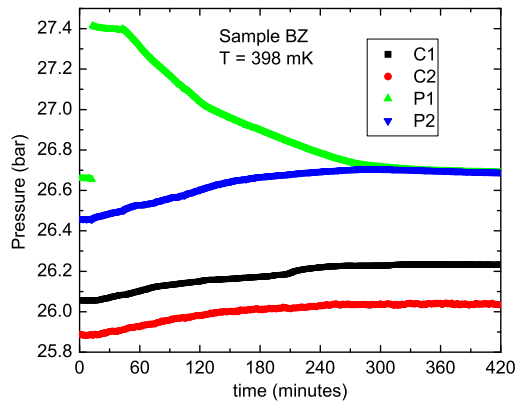}}
\caption{\label{fig:figureBZ} (color online) Sample BZ, which
showed the more normal behavior of flow to equilibration, P1 = P2.
Sample BZ, is the same sample as BY, designated BZ with an
injection following the completion of the behavior seen in figure
\ref{fig:figureBY}. }
\end{figure}

While most of our data sets that showed clear evidence for flow
behaved in a common way (P2 increased linearly with time and the
pressures recorded on the cell capacitors increased), and
occasionally we saw changes in the cell pressure with more limited
changes in P2, we did see one rather more unusual data set. A
sample was grown fresh as sample BV and flowed nicely at 399 mK.
It was then injected again (sample BW) and showed no evidence for
flow. It was then subjected to a withdrawal (sample BX) and again
showed no evidence for flow, an unusual event for a withdrawal. It
was then subjected to a further injection (sample BY) and showed
evidence for flow that then stopped prior to equilibration, figure
\ref{fig:figureBY}. A subsequent injection (sample BZ) resulted in
a normal flow to equilibration, figure \ref{fig:figureBZ}. An
additional injection (sample CA) showed typical further evidence
for flow. There were few anomalous situations such as described by
the sequence BV - CA, and we simply note this one here to be
complete.

\section{Tables of Sample Characteristics}

Here we list each of the samples of solid helium that we have
studied (Table I, II), and each of the sets of samples for which
we have done thermal cycling sequences (Table III). In tables I
and II we list each sample by the code letter that was assigned to
it. Samples were created or studied in the order of the code
letters in almost all cases. Note that a new code letter was given
to a sample each time a change was made in the sample. So, for
example, sample A was grown fresh from superfluid, so was sample
B, but sample C was the code given to sample B after it had been
warmed to 1.25 K. Missing code letters indicate samples that were
useless or untrustworthy for one reason or another (e.g.,
temperature instabilities in the apparatus, a helium transfer
mid-run, etc.).

\begin{table*}
\caption{\label{tab:samp} Table of flow measurements made for the
solid Helium flow experiments showing: (1) sample - each
measurement was deemed a separate sample (2) run - each run was a
chain of samples all from the same growth (3) history - either how
the sample was grown, or what sample it came from (4) sample
Temperature (5) sample pressure (6) initial pressure step put into
line 1 (7) flow rate observed in line 2.  a) Pressure added to
line 2.  b) A first addition of 0.41 bar showed no flow over 6.5
hours. c) Subtraction from line 4.  d) A number of small pressure
steps (0.2 bar). }
\begin{ruledtabular}
\begin{tabular}{ccccccc}
sample & run & history & T$_{sample}$(K) & P$_{sample} $(bar) & $\Delta P_i $(bar) & dP2/dt (mbar/s)\\
\hline
A & sf1-3a\_sh1 & superfluid & 0.398 & 26.75 & 1.482 & 0.0051 \\
B & sf1-3a\_sh2 & superfluid & 0.386 & 26.41 & 1.752 & 0.0290 \\
C & sf1-3a\_sh2 & B & 1.25  &  25.65 & 2.432 & 0.0000 \\
D & sf1-3\_sh5  & Blocked Capillary & 0.285 & 28.95 & 1.705 & 0.0000 \\
E & sf1-3\_sh5  & D & 0.900 & 28.82 & 1.775 & 0.0000 \\
F & sf1-3\_sh4  & Blocked Capillary & 0.380 & 28.70 & 2.156 & 0.0000\\
G & sf1-3b\_sh1 & superfluid & 0.380 & 26.68 & 1.779 & 0.0027 \\
H & sf1-3b\_sh1 & G & 0.444 & 27.30 & 0.778 & 0.0000 \\
J & sf1-3b\_sh1 & H & 0.460 & 27.24 & 1.797 & 0.0000\\
K & sf1-3b\_sh1 & J & 0.420 & 27.16 & -2.233 & -0.0127 \\
L & sf1-4b\_sh2 & superfluid & 0.391 & 26.89 & -1.221 & 0.0472 \\
M & sf1-3b\_sh2 & superfluid & 0.392 & 26.36 & 0.950 & 0.0121\\
N & sf1-3b\_sh2 & M & 0.804 & 26.54 & 1.032 & 0.0000 \\
O & sf1-3b\_sh2 & N & 0.384 & 26.52 & 1.005 & 0.0000 \\
P & sf1-3b\_sh2 & O & 0.384 & 26.52 & -2.300 & -0.0804 \\
Q & sf1-3b\_sh2 & O & 0.400 & 25.93 & 0.616 & 0.0067\\
R & sf1-3b\_sh2 & Q & 0.396 & 26.13 & 0.801 & 0.0075 \\
S & sf1-3b\_sh3 & Blocked Capillary & 0.400 & 27.25 & -2.718c & -0.0667 \\
T & sf1-3b\_sh3 & Blocked Capillary & 0.392 & 26.51 & 0.691 a & 0.0171 \\
U & sf1-3b\_sh4 & Blocked Capillary & 0.380 & 26.54 & 0.541 a & 0.0120 \\
V & sf1-3b\_sh5 & Blocked Capillary & 0.395 & 26.13 & 0.844 & 0.0070 \\
W & sf1-3b\_sh6 & Blocked Capillary & 0.290 & 29.48 & 0.582 & 0.0000 \\
X & sf1-3b\_sh7 & Superfluid & 0.393 & 26.13 & 1.062 b & 0.0134 \\
Y & sf1-3b\_sh7 & X & 0.391 & 26.43 & 0.462 & 0.0093 \\
Z & sf1-3b\_sh7 & Y & 0.818 & 26.83 & 0.465 & 0.0000 \\
AA & sf1-3b\_sh7 & Z & 0.388 & 26.79 & 0.457 & 0.0000 \\
AB & sf1-3b\_sh8 & superfluid & 0.398 & 26.75 & 0.430 & 0.0230 \\
AC & sf1-3b\_sh8 & AB & 1.02 & 25.93 & 0.518 & 0.0000 \\
AD & sf1-3b\_sh9 & Superfluid & 0.800 & 25.93 & 0.483 & 0.0000 \\
AE & sf1-3b\_sh10 & superfluid & 0.850 & 25.93 & d & 0.0000 \\
%AF & sf1-3c\_sh1 & superfluid & ??? & ??? & ??? & ??? \\
AG & sf1-3c\_sh1 & AF & 0.363 & 25.95 & 0.414 & 0.0195 \\
AH & sf1-3c\_sh2 & superfluid & 0.363 & 25.92 & 0.723 & 0.0434 \\
AJ & sf1-3c\_sh4 & superfluid & 0.498 & 26.18 & 0.633& 0.0111 \\
AK & sf1-3c\_sh4 & AJ & 0.498 & 26.37 & 0.443& 0.0026 \\
AL & sf1-3c\_sh4 & AK & 0.359 & 26.52 & 0.818& 0.0076 \\
AM & sf1-3c\_sh4 & AL & 0.498 & 26.69 & 0.423 & 0.0000 \\
AN & sf1-3c\_sh5 & superfluid & 0.359 & 26.30 & 0.473 & 0.0203\\
AO & sf1-3c\_sh6 & superfluid & 0.358 & 26.51 & 0.451 & 0.0086\\
AP & sf1-3c\_sh6 & AO & 0.608 & 26.61 & 0.421 & 0.0000 \\
AQ & sf1-3c\_sh6 & AP & 0.358 & 26.62 & 0.350 & 0.0000 \\
AR & sf1-3c\_sh7 & superfluid & 0.359 & 26.25 & 0.391 & 0.0088 \\
AS & sf1-3c\_sh7 & AR & 0.608 & 26.37 & 0.531 & 0.0000 \\
AT & sf1-3c\_sh7 & AS & 0.359 & 26.36 & 0.548 & 0.0000 \\
AU & sf1-3c\_sh7 & AT & 0.359 & 26.37 & -1.307 & -0.0302 \\
AV & sf1-3c\_sh7 & AU & 0.358 & 26.39 & 0.662 & 0.0054 \\
AW & sf1-3c\_sh8 & superfluid & 0.608 & 25.81 & 0.500 & 0.0000\\
AX & sf1-3c\_sh8 & AW & 0.600 & 25.82 & -0.793 & 0.0000 \\
AY & sf1-3c\_sh8 & AX & 0.360 & 25.78 & 0.489 & 0.0000 \\
BA & sf1-3c\_sh9 & superfluid & 0.360 & 26.37 & 0.538 & 0.0043 \\
BC & sf1-4\_sh1 & superfluid & 0.400 & 25.91 & 0.606 & 0.0209\\
BD & sf1-4\_sh1 & BC & 0.390 & 25.86 & 0.664 & 0.0146\\
BE & sf1-4\_sh1 & BD & 0.498 & 26.03 & 0.542 & 0.0000\\
BF & sf1-4\_sh1 & BE & 0.394 & 26.10 & 0.394 & 0.0364\\

\end{tabular}
\end{ruledtabular}
\end{table*}

\begin{table*}
\caption{ Continuation of Table I. }
\begin{ruledtabular}
\begin{tabular}{ccccccc}
sample & run & history & T$_{sample}$(K) & P$_{sample} $(bar) & $\Delta P_i $(bar) & dP2/dt (mbar/s)\\
\hline
BJ & sf1-4\_sh3 & superfluid & 0.397 & 26.45 & 0.449 & 0.0112\\
BK & sf1-4\_sh3 & BJ & 0.452 & 26.52 & 0.432 & 0.0000\\
BL & sf1-4\_sh3 & BK & 0.397 & 26.58 & 0.398 & 0.0000\\
BM & sf1-4\_sh3 & BL & 0.396 & 26.50 & 0.277 & 0.0000\\
BO & sf1-4\_sh4 & superfluid & 0.398 & 26.03 & 0.427 & 0.0095\\
BP & sf1-4\_sh5 & superfluid & 0.400 & 26.23 & 0.418 & 0.0140\\
BQ & sf1-4\_sh5 & BP & 0.449 & 26.53 & 0.475 & 0.0058\\
BR & sf1-4\_sh5 & BQ & 0.400 & 26.52 & 0.416 & 0.0039\\
BS & sf1-4\_sh6 & superfluid & 0.400 & 26.55 & 0.414 & 0.0065\\
BT & sf1-4\_sh6 & BS & 0.547 & 26.49 & 0.419 & 0.0000\\
BU & sf1-4\_sh6 & BT & 0.398 & 26.50 & 0.515 & 0.0052\\
BV & sf1-4\_sh7 & superfluid & 0.399 & 26.17 & 0.558 & 0.0117\\
BW & sf1-4\_sh7 & BV & 0.398 & 26.18 & 0.918 & 0.0000 \\
BX & sf1-4\_sh7 & BW & 0.398 & 26.18 & -1.122 & 0.0000 \\
BY & sf1-4\_sh7 & BX & 0.398 & 26.01 & 0.437 & 0.0178 \\
BZ & sf1-4\_sh7 & BY & 0.398 & 26.06 & 0.748 & 0.0215 \\
CA & sf1-4\_sh7 & BZ & 0.397 & 26.23 & 0.573 & 0.0145 \\

\end{tabular}
\end{ruledtabular}
\end{table*}

Table III is a tabulation of sets of samples that were part of a
sequence of measurements that began with the sample with code
letter indicated in the first column. So, for example, the first
entry represents a sequence that began with sample M, which was
created from superfluid at 392 mK. Sample M was then warmed to 804
mK (and denoted sample N, Table I). Following measurement at 804
mK, sample N was cooled to 384, renamed sample O, and studied.
Table II shows such triples.  Note also that for some of these
sequences of three, like M, N, O, a continuation of the sequences
(e.g. P, Q, R) took place, which can be seen in Table I.

\begin{table*}
\caption{\label{tab:cycle} Thermal cycling of solid helium samples
and its effect on flow through the sample.  A given series begins
the the sample denoted in parentheses in column 1.   Temperatures
and pressures are those of the cell. In all cases atoms were added
to line 1 and dP2/dt is the measured rate of change of P2.  a) No
change in the pressure of P2 was recorded for 60 minutes followed
by rapid pressure relaxation.  b) No change in P2 was recorded,
however the cell pressure increased and P1 decreased.}
\begin{ruledtabular}
\begin{tabular}{cccccccccc}
Series & T$_A$ (mK) & P$_A $(bar) & dP2$_A$/dt (mbar/s) & T$_B$
(mK)
& P$_B$ (bar) & dP2$_B$/dt (mbar/s) & T$_C$ (mK) & P$_C$ (bar) & dP2$_C$/dt (mbar/s) \\
\hline
1  (M) & 392 & 26.36 & 0.0121 & 805 & 26.54 & 0.0000  & 385 & 26.52 & 0.0000  \\
2  (Y) & 391 & 26.43 & 0.0093 & 800 & 26.83 & 0.0000  & 388 & 26.79 & 0.0000  \\
3 (AK) & 498 & 26.37 & 0.0026 & 360 & 26.52 & 0.0076  & 498 & 26.69 & 0.0000 b\\
4 (AO) & 358 & 26.51 & 0.0086 & 608 & 26.61 & 0.0000  & 358 & 26.62 & 0.0000  \\
5 (AR) & 359 & 26.25 & 0.0088 & 608 & 26.37 & 0.0000  & 357 & 26.36 & 0.0000  \\
6 (BD) & 390 & 25.86 & 0.0146 & 498 & 26.03 & 0.0000 b & 394 & 26.10 & 0.0364 a\\
7 (BJ) & 397 & 26.45 & 0.0112 & 452 & 26.52 & 0.0000 b & 397 & 26.58 & 0.0000 b\\
8 (BP) & 400 & 26.23 & 0.0140 & 449 & 26.53 & 0.0058  & 400 & 25.52 & 0.0039  \\
9 (BS) & 400 & 26.55 & 0.0065 & 547 & 26.49 & 0.0000  & 398 &
26.50 & 0.0052
\end{tabular}
\end{ruledtabular}
\end{table*}

%Included for Gather Purpose only:
%input "C:\localtexmf\bibtex\bib\ref.bib"
\clearpage
\bibliography{ref}% Produces the bibliography via BibTeX.
\end{document}